\documentclass[conference]{IEEEtran}
\IEEEoverridecommandlockouts
\usepackage{svg}
\usepackage{cite}
\usepackage{amsmath,amssymb,amsfonts}
\usepackage{algorithmic}
\usepackage{graphicx}
\usepackage{textcomp}
\usepackage{xcolor}
\usepackage{url}

\def\BibTeX{{\rm B\kern-.05em{\sc i\kern-.025em b}\kern-.08em
    T\kern-.1667em\lower.7ex\hbox{E}\kern-.125emX}}
\begin{document}

\title{Towards practicable Machine Learning development using AI Engineering Blueprints \\
}

\makeatletter
\newcommand{\linebreakand}{%
\end{@IEEEauthorhalign}
\hfill\mbox{}\par
\mbox{}\hfill\begin{@IEEEauthorhalign}
}
\makeatother

\author{\IEEEauthorblockN{Nicolas Weeger}
\IEEEauthorblockA{\textit{Ansbach UAS}\\
Ansbach, Germany\\
nicolas.weeger@hs-ansbach.de}
\and
\IEEEauthorblockN{ Annika Stiehl}
\IEEEauthorblockA{\textit{Ansbach UAS}\\
	Ansbach, Germany\\
	annika.stiehl@hs-ansbach.de}
\and
\IEEEauthorblockN{Jóakim von Kistowski}
\IEEEauthorblockA{\textit{Aschaffenburg UAS}\\
	Aschaffenburg, Germany \\
	joakim.vonkistowski@th-ab.de}
\linebreakand 
\IEEEauthorblockN{Stefan Geißelsöder}
\IEEEauthorblockA{\textit{Ansbach UAS}\\
	Ansbach, Germany \\
	stefan.geisselsoeder@hs-ansbach.de}
\and
\IEEEauthorblockN{Christian Uhl}
\IEEEauthorblockA{\textit{Ansbach UAS}\\
	Ansbach, Germany \\
	christian.uhl@hs-ansbach.de}

}

\maketitle

\begin{abstract}
The implementation of artificial intelligence (AI) in business applications holds considerable promise for significant improvements. 
The development of AI systems is becoming increasingly complex, thereby underscoring the growing importance of AI engineering and MLOps techniques. 
Small and medium-sized enterprises (SMEs) face considerable challenges when implementing AI in their products or processes. These enterprises often lack the necessary resources and expertise to develop, deploy, and operate AI systems that are tailored to address their specific problems.  

Given the lack of studies on the application of AI engineering practices, particularly in the context of SMEs, this paper proposes a research plan designed to develop blueprints for the creation of proprietary machine learning (ML) models using AI engineering and MLOps practices. These blueprints enable SMEs to develop, deploy, and operate AI systems by providing reference architectures and suitable automation approaches for different types of ML.

The efficacy of the blueprints is assessed through their application to a series of field projects. This process gives rise to further requirements and additional development loops for the purpose of generalization. The benefits of using the blueprints for organizations are demonstrated by observing the process of developing ML models and by conducting interviews with the developers.

\end{abstract}


\section{Introduction}
Artificial intelligence (AI) is transforming numerous industries and application domains. It is crucial for organizations to adopt AI techniques in order to achieve business success  \cite{enholmArtificialIntelligenceBusiness2022, loureiroArtificialIntelligenceBusiness2021}. 
SMEs in particular can benefit significantly from adopting AI techniques. They can improve or even establish capabilities in certain areas, such as customer experience, production monitoring, and decision-making processes \cite{bhaleraoStudyBarriersBenefits2022}. 

The proprietary development of ML models for use as or within a product, called AI systems, in an organizational context can lead to a number of challenges \cite{fischerAISystemEngineering2020,lwakatareDevOpsAIChallenges2020,schonbergerArtificialIntelligenceSmall2023, nascimentoUnderstandingDevelopmentProcess2019}. This includes  understanding of the intricacies of AI, including its functional requirements and operational scenarios. The integration of additional processes, such as data generation and preprocessing or model training and deployment, with the traditional software engineering process can potentially create organizational constraints, particularly for SMEs \cite{schonbergerArtificialIntelligenceSmall2023}. The ML model development lifecycle includes additional practices beyond DevOps for the data and models. These include MLOps and DataOps techniques that provide a culture, practices, and tools for handling data and models. In addition, the system architecture for AI systems must be aligned with the requirements of the underlying model. Training and inference environments, along with data storage and versioning of the different artifacts, must be incorporated in order to enable the system to function as an AI system. 

Consequently, the effective implementation of AI systems requires a well-designed architecture that is tailored to the specific requirements of the intended AI application. This paper discusses the importance of AI engineering practices and the concept of developing blueprints tailored to the requirements of different types of AI and stages of development. 

We will discuss the content of these blueprints and how they can be utilized by organizations. The discussion will commence with a preliminary result in the form of a pipeline and its sub-pipelines.

The results of this research will assist organizations to address these challenges and streamline the development, deployment and operation of AI systems. Consequently, they will be enabled to adopt the blueprints by providing reference architectures and suitable automation approaches for various types of AI.

\section{Background and Related Work}

\subsection{AI Engineering}

AI engineering is an evolution of the field of software engineering and, due to the rapid growth of ML developments, it is an emerging field. AI engineering is currently at the maximum ``peak of inflated expectations", as evidenced by Gartner's AI Hype Cycle for 2024\footnote{\url{https://www.gartner.com/en/articles/hype-cycle-for-artificial-intelligence}}. According to Gartner, ``AI engineering is the foundation for enterprise delivery of AI and GenAI at scale. Most organizations lack the data, analytics and software foundations to move individual AI projects to production at scale - much less operate a portfolio of AI solutions at scale."
Over a dozen projects were examined in \cite{boschEngineeringAISystems2021}, where AI engineering challenges led to problems with productizing ML models. The study found that the majority of companies that develop machine learning models encounter difficulties when trying to move them into production. They provide a research agenda and overview of the issues that need to be addressed in this direction. \cite{groteCaseStudyAI2023} point out that while the field of software engineering research has been extensively discussed, AI engineering has been much less addressed. Only a limited number of publications present concrete experiences related to the application of AI engineering principles. They selected ten AI engineering practices in several categories from the literature and applied them to an example implementation to evaluate the practices and their systems architecture. 

Furthermore, discussions and questionnaires, especially with SMEs, have shown that there is a desire to implement AI in their systems. However, the success of their model implementation and productization depends on overcoming the challenges mentioned above. Thus, applying AI engineering practices can help organizations streamline the development, deployment and operation of machine learning models.

\subsection{MLOps}

The idea of MLOps is to provide techniques and tools for the deployment and operation of AI systems \cite{symeonidisMLOpsDefinitionsTools2022}. The goal is to devise a strategy for solving real-world problems with the deployment of ML models. Several studies review various literature in this area and offer pipelines, taxonomies, tools, methods and challenges in this area \cite{testiMLOpsTaxonomyMethodology2022,kreuzbergerMachineLearningOperations2023,steidlPipelineContinuousDevelopment2023}. 

\cite{najafabadiAnalysisMLOpsArchitectures2024} conduct a systematic mapping study for MLOps architectures and point out 35 architecture components, describe different architecture variants for different use cases and provide popular tools to use for these architecture components. 

\subsection{Other Related Work}
In \cite{wostmannConceptionReferenceArchitecture2020} a reference architecture for the specific use cases in the process industry dealing with edge devices was presented. They demonstrated the architecture by implementing a case study for a real-world use case and proved the functionality with this application. 

\cite{paakkonenExtendingReferenceArchitecture2020} developed a reference architecture to facilitate the use case of big data in edge computing ML techniques. They come up with different views on the architecture of model development and deployment for this specific use case. 

Another study \cite{vandenheuvelModelDrivenMLOpsIntelligent2020} presents a vision for ``disciplined, repeatable and transparent model-driven development and Machine-Learning operations (MLOps) of intelligent enterprise applications." They provide a three-stage metamodel for model based development of AI/ML blueprints based on intelligent application architecture.

With scope on software and architecture, design patterns for AI based systems are discussed in several studies \cite{heilandDesignPatternsAIbased2023,sharmaDesignPatternsMachine2019,cabralInvestigatingImpactSOLID2024,takeSoftwareDesignPatterns2021}. They provide an overview of design patterns, adapted or newly generated for AI use cases, and show the application and the resulting benefits of using them to develop machine learning models.  

In summary, the literature provides insights into the importance, potential architectures, and principles for AI engineering and MLOps practices. However, the application of these insights is currently focused on a few reference architectures in specific domains, such as big data or edge devices. Other studies focus on defining architectures and patterns and prove their applicability in case studies.

The principles of AI engineering provide the foundation for the development of the blueprints proposed in this paper. MLOps pipelines and tools as well as existing reference architectures and frameworks are employed to support the development of AI systems, thereby streamlining, standardizing, and accelerating the process. 
Software and architecture design patterns are used to describe the development in order to fulfill the non-functional requirements (NFRs) for the different blueprints.
The application in field projects enables flexible, highly automated deployment and resource-efficient operation for different requirements in SMEs.

\section{Methodology}

Since the literature is sparse when it comes to practical examples of AI engineering directly usable for SMEs, this work is intended to develop blueprints for common AI use cases. The blueprints comprise the description of suitable reference architectures and reference applications, automation pipelines, and tools for model development, deployment, and operation.
 
The extant literature offers a variety of interpretations of MLOps pipelines, which encompass the principles of DataOps and DevOps as they relate to ML \cite{testiMLOpsTaxonomyMethodology2022,kreuzbergerMachineLearningOperations2023,steidlPipelineContinuousDevelopment2023}. Previous considerations of these MLOps pipelines have led to the formulation of a guiding pipeline for the development of blueprints (Fig. \ref{overall_pipeline}). 

\begin{figure}[htbp]
	\centering
	\includegraphics[width=\columnwidth]{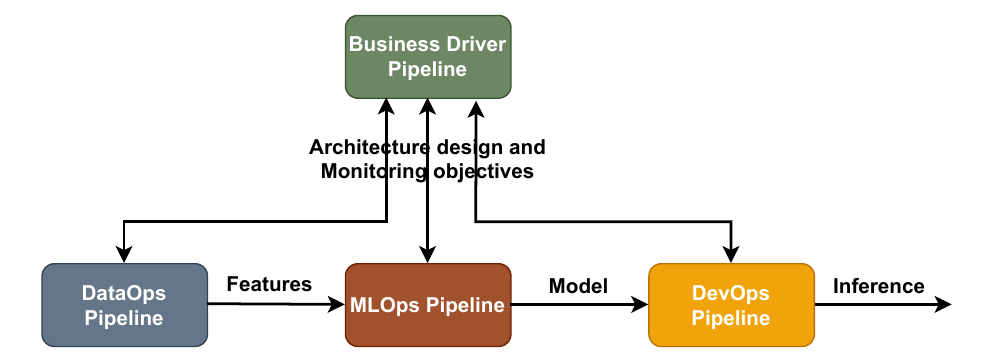}
	\caption{Guiding pipeline with four sub-pipelines.}
	\label{overall_pipeline}
\end{figure}

First, the business requirements are defined and the architectural components are specified. This serves as a framework for implementing and validating all subsequent stages. The DataOps pipeline prepares the data for model training in the MLOps pipeline. The resulting model artifact is then incorporated into software using the DevOps pipeline to enable inference. 

The following subsections provide a detailed breakdown of these sub-pipelines. They elaborate the methodology for developing blueprints for each of these pipelines, for different types of AI, and deployment scenarios. The AI types, including computer vision, time series analysis, reinforcement learning, and generative AI, as well as the manifold deployment scenarios possess systematically different requirements that lead to varied architecture definitions and thus the necessity for multiple blueprints. 

The combination of these blueprints according to the requirements of a project will facilitate a standardized and simplified approach that will enhance the efficiency of model development and operation.

\subsection{Business driver pipeline}
\begin{figure}[htbp]
	\centering
	\includegraphics[width=\columnwidth]{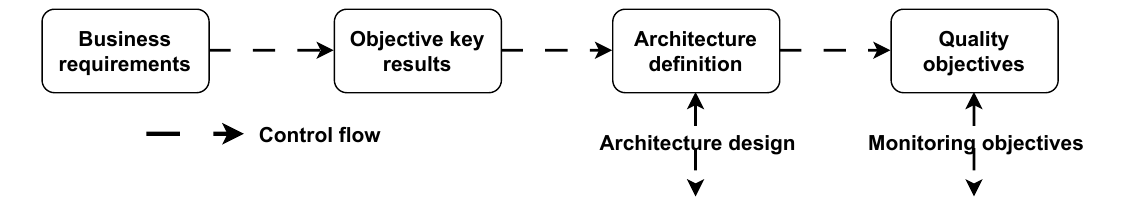}
	\caption{Business driver pipeline for  defining requirements, architecture and objectives.}
	\label{business_driver_pipeline}
\end{figure}

The business driver pipeline (Fig. \ref{business_driver_pipeline}) provides a methodology for defining requirements, examining measurable objective key results (OKRs), and defining architecture for the software and ML components as well as quality objectives for the monitoring.
The business requirements are then discussed with the organizational stakeholders and subsequently combined with the NFRs of the corresponding AI type and translated into measurable and testable OKRs. Subsequently, these are employed to specify the architectural framework for the software components, including the different levels of design patterns and tools to be used for the pipeline architecture, in addition to the model architecture and data structure alternatives. Objectives for monitoring the model performance, environment utilization and data quality are defined to allow the generation of monitoring alerts in production scenarios.

\subsection{DataOps pipeline}

\begin{figure}[htbp]
	\centering
	\includegraphics[width=\columnwidth]{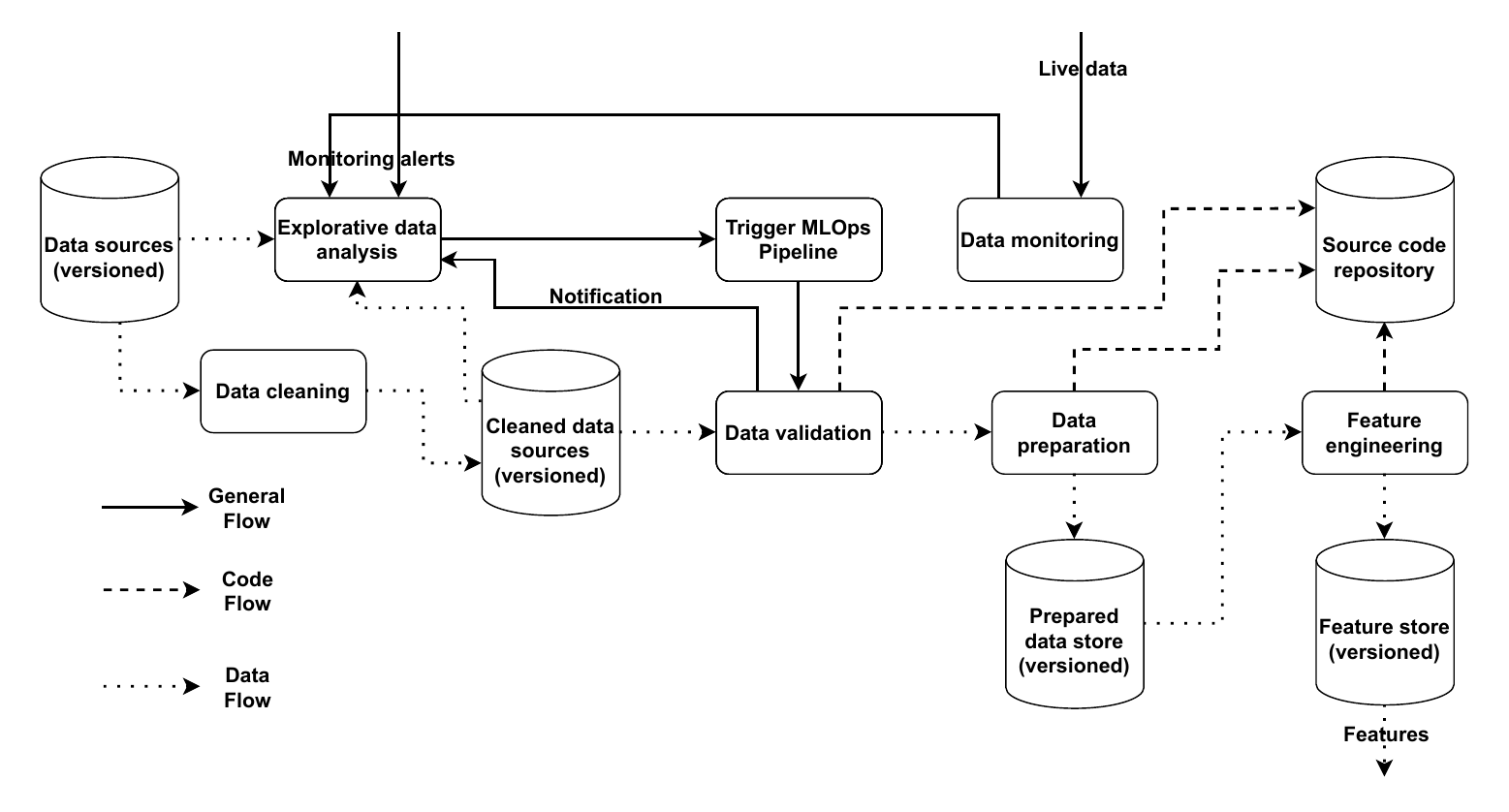}
	\caption{DataOps pipeline to prepare the data for model training.}
	\label{dataops_pipeline}
\end{figure}

The DataOps pipeline (Fig. \ref{dataops_pipeline}) includes all stages of data processing for model training. This covers the full range of activities, from exploratory data analysis and data cleaning to data validation, preparation and feature engineering. A key aspect of the DataOps pipeline is the versioning of the data. This enables the reproducibility of each stage of data manipulation. Note that not all use cases will require all of these steps. The blueprints should include options for tools that can be adapted to different use cases, as well as predefined steps that can be used for data processing with different requirements.

\subsection{MLOps pipeline}

\begin{figure}[htbp]
	\centering
	\includegraphics[width=\columnwidth]{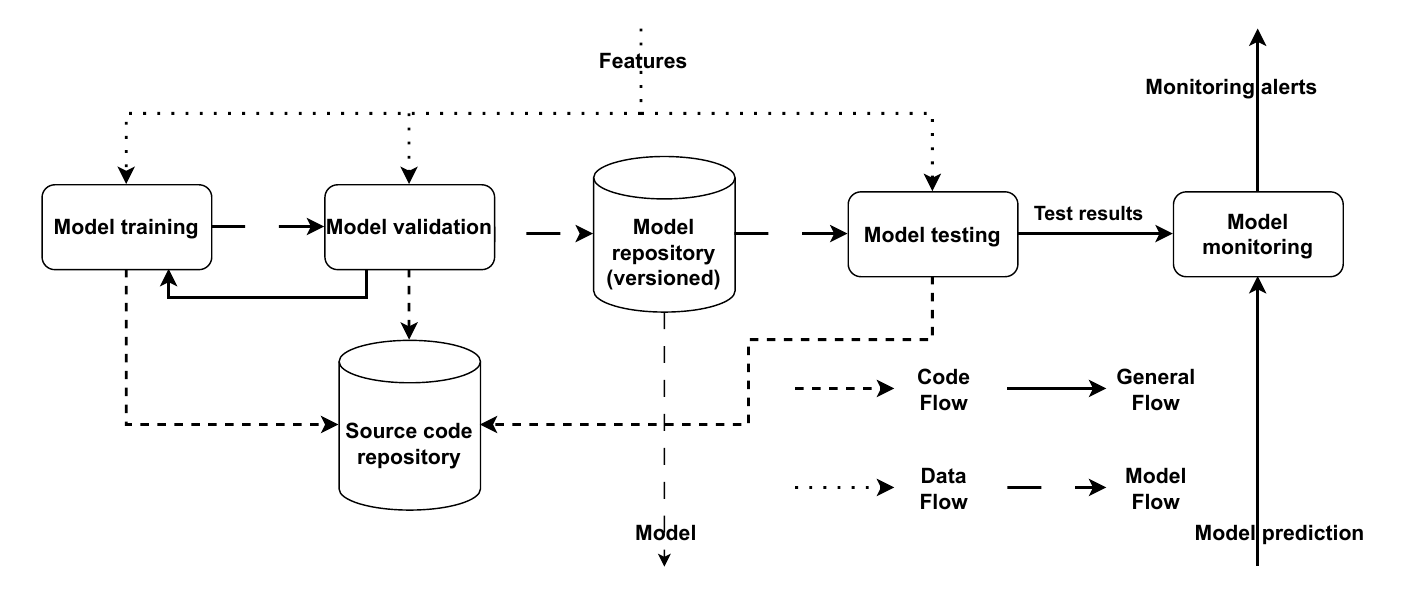}
	\caption{MLOps pipeline for model training, validation and versioning.}
	\label{mlops_pipeline}
\end{figure}

The MLOps pipeline (Fig. \ref{mlops_pipeline}) incorporates components for model training, validation, and testing.  These components utilize features from the feature store that was created by the DataOps pipeline. The blueprints for this phase include information about the tools to be employed for the automation of the training process, the tracking of experiments, and the storage of versioned model artifacts for the experiments. Furthermore, the blueprints should identify different training environments that facilitate the implementation of high-performance training for different types of ML and data.

\subsection{DevOps pipeline}

\begin{figure}[htbp]
	\centering
	\includegraphics[width=\columnwidth]{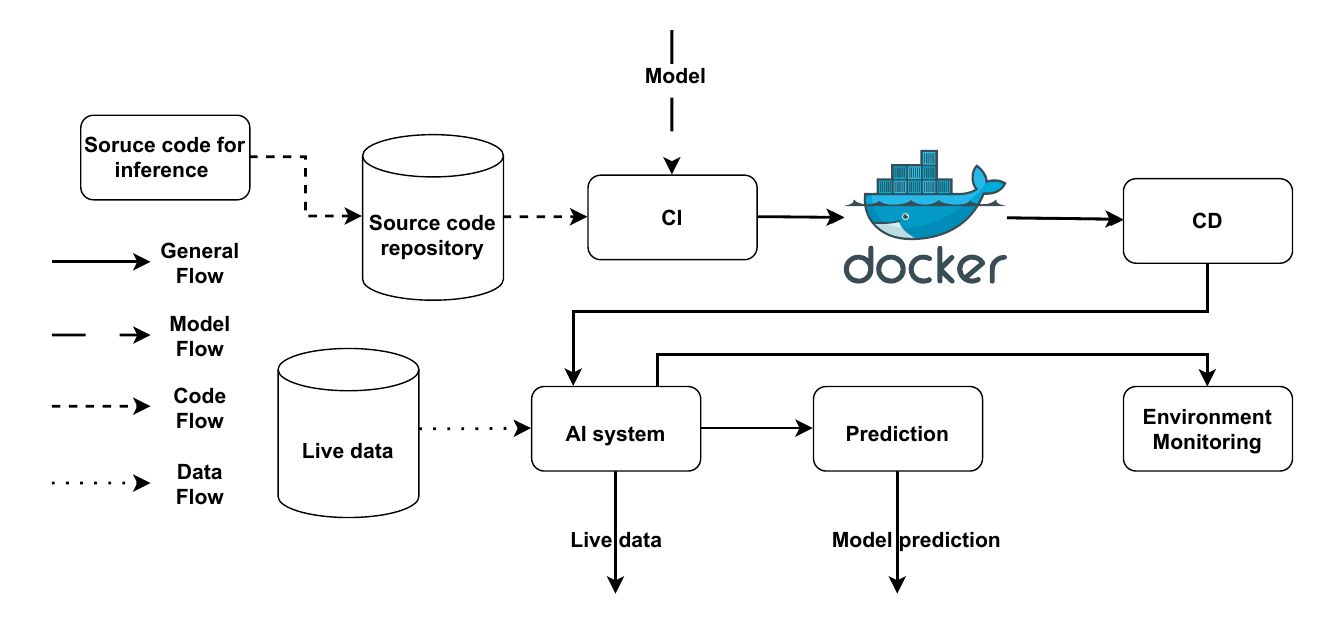}
	\caption{DevOps pipeline for system integration of the model artifact.}
	\label{devops_pipeline}
\end{figure}

The DevOps pipeline (Fig. \ref{devops_pipeline}) covers the system integration of the final model artifact, using continuous integration (CI) and continuous deployment (CD) for the prediction on live data in an AI system. This includes source code to enable inference, such as batch and API interfaces, and the packaging or containerizing of the usable model service for productization. The blueprints include options for interface implementation and different deployment strategies, such as cloud, edge or standalone deployment, depending on the requirements of the use case. Monitoring of model performance, environment utilization, and live data quality is also defined and discussed.

\subsection{Research method}
In order to examine the blueprints and validate their usability for enterprises, the design science research (DSR) method \cite{hevnerDesignScienceInformation2004a, ivarssonMethodEvaluatingRigor2011}, shall be employed. By means of interviews and a comprehensive review of relevant literature, the challenges and requirements of SMEs are identified, and potential avenues for improvement are ascertained.

Based on these findings, business requirements can be established in collaboration with the relevant stakeholders, aligning with their specific needs. By integrating the business requirements with the requirements of the various types of AI being utilized, for instance, algorithms, data storage, computational demands, and NFRs, a comprehensive framework can be devised to facilitate the verification of the designed artifacts. Subsequently, these can be subjected to iterative testing and validation. Finally, the artifacts can be deployed in the stakeholders' projects as a field test. This process is then to be repeated until the requirements have been finalized and the artifacts have been demonstrated to fulfill them. The process is repeated in multiple projects  to generalize the findings, making them as applicable as possible for SMEs.

\section{Conclusion}
This paper shows the need for research that helps small and medium sized enterprises to integrate development of ML models into their organizations. To address the challenges inherent in this process, the proposal suggests the use of blueprints that include reference architectures, pipelines, and tools for model development, deployment, and operation, in conjunction with reference applications tailored to the specific requirements of different types of AI.

It is essential that the blueprints be as specific as possible with regard to a particular type of AI and deployment strategy, while also being sufficiently general to be beneficial to a wide range of applications. 
To truly enable the streamlined application of AI for SMEs, we argue that requirements from different projects must be successively integrated into generalized blueprints. In turn, these blueprints must be tested against multiple use-cases in real world applications.

\bibliographystyle{IEEEtran} 
\bibliography{IEEEabrv,ML_blueprints_bibliography.bib}

\end{document}